\documentclass[12pt,a4paper]{article}

\newcommand{\be}{\begin{equation}}
\newcommand{\ee}{\end{equation}}
\newcommand{\bea}{\begin{eqnarray}}
\newcommand{\eea}{\end{eqnarray}}

\begin{document}

\title{A Simple Signal of Noncommutative Space}

\author{Ciprian Sorin Acatrinei\thanks{On leave from: {\it National Institute of
        Nuclear Physics and Engineering -
        P.O. Box MG-6, 76900 Bucharest, Romania}; e-mail:
        acatrine@physics.uoc.gr.} \\
        Department of Physics, University of Crete, \\
        P.O. Box 2208, Heraklion, Greece}

\date{November 12, 2003}

\maketitle

\begin{abstract}
We propose a simple low-energy classical experiment
in which the effects of noncommutativity can be clearly separated 
from commutative physics. The ensuing bound on the noncommutative scale 
is remarkable, especially in view of its elementary derivation. 
\end{abstract}

Noncommutative mechanics has recently been under intensive investigation \cite{ncqm,bounds1}.
It uses the following basic commutators for the coordinates $q_i$ and momenta $p_i$, 
\be
[q_i,q_j]=i\theta_{ij}, \qquad [q_i,p_j]=i\delta_{ij}, \qquad [p_i,p_j]=iF_{ij}, \label{cr1}
\ee 
which are a generalization of the Heisenberg commutation relations. 
The time evolution of an operator  follows from its commutator with the prescribed Hamiltonian.
In the classical case the commutators in (\ref{cr1}) are replaced by Poisson brackets,
\be
\{q_i,q_j\}=\theta_{ij}, \qquad \{q_i,p_j\}=\delta_{ij}, \qquad\{p_i,p_j\}=F_{ij}.   \label{pb}
\ee 
Above, $\delta_{ij}$ is a unit matrix, 
whereas $F_{ij}$ and $\theta_{ij}$ 
are
functions of both the coordinates and the momenta, 
to be restricted only by the Jacobi identities.
This note will consider the case in which $\theta$ is a nonvanishing constant diagonal matrix, 
whereas a generic $F(q_i,p_j)$ is allowed. A dynamical effect due to the gradient of $F$ will be reported.
For simplicity in presentation we will work in two dimensions, $i,j=1,2$ in (\ref{pb}), although our considerations
extend straightforwardly to any space dimensionality.
Lower case latin indices will always take two values, e.g. $i,j,k,s,t=1,2$, 
and no distinction will be necessary between up and down indices in the following.
In two dimensions the only nontrivial Jacobi identities read, 
with the notation $\theta_{12}=-\theta_{21}\equiv\theta$,
\be
\partial_{p_1}F=-\theta\partial_{q_2}F, \qquad \qquad \partial_{p_2}F=\theta\partial_{q_1}F. \label{jacobi}
\ee
The above relations 
ensure the invariance of Eqs. (\ref{pb}) under time evolution.
Once we disregard the constant $F$ solution
\footnote{Experimentally $F$ is never constant throughout all space, moreover
the effect we are after arises when $\nabla F\neq 0$.},
(\ref{jacobi}) implies that $F$ is a function of the combinations $\bar{q}_1=q_1+\theta p_2$ and $\bar{q}_2=q_2-\theta p_1$,
in brief $F=F(\bar{q}_1,\bar{q}_2)=F(\bar{q}_i)=F(q_i+\theta_{ij}p_j)$.
If $\theta$ is then assumed to be small, which is the maximum the real world allows for, one has
\be
\left.
F(\bar{q}_i)=F(q_i+\theta_{ij}p_j)=F(q_i)+\theta_{st}p_t
\partial_{\bar{q}_s}F_{ij}(\bar{q})
\right|_{\bar{q}_i=q_i}.
\label{F}
\ee 
To see the consequences of such an expression, consider the equations of motion for a free Hamiltonian
$H=\frac{p_i^2}{2m}$,
\be
m\dot{q_i}=m\{q_i,H\}=p_i, \qquad m\dot{p_i}=m\{p_i,H\}=F_{ij}(q_s+\theta_{st}p_t)p_j , \label{em}
\ee
where $\frac{df}{dt}\equiv\dot{f}$.
If $\theta=0$, $F_{ij}$ just provides a magnetic field of intensity $F_{ij}(q)$.
If $\theta$ is small, (\ref{F}) implies that
\be
m\ddot{q}_i=F(q^s+\theta_{st}p_t)_{ij}\dot{q}_j\simeq 
F(q^s)_{ij}\dot{q}_j+m\partial_s F_{ij}(q) \theta_{st}\dot{q}_t\dot{q}_j \label{test1}.
\ee
Eq. (\ref{test1}) displays the superposition of a usual electromagnetic background, linear in the velocity,
and of a force proportional to the mass and
quadratic in the velocities, $\partial_s F(q)_{mj} \theta_{st}\dot{q}_t\dot{q}_j$ -- echoing gravity.
However $\gamma^s_{tj}\equiv \partial_s F(q)_{mj} \theta_{st}$ simulates a gravitational connection
only in a particular reference frame, since 
$\gamma^s_{tj}$ does not behave like a Christoffel symbol under generalized coordinate transformations.

Experimentally speaking, it is important that a nonzero $\theta$ {\em requires} 
a nonconstant $F$ to be a function of both position and velocity, $F(q,p)$.
This suggests that noncommutativity could be seen through the effects of the additional term in (\ref{test1}).
A simple experiment is actually possible and provides - given its elementary character -
a remarkably good bound for $\theta$.

Consider the 2-dimensional case of motion on a plane labelled by coordinates $q_1=x$, $q_2=y$,
with velocities $v_x=\dot{x}=p_1/m$, $v_y=\dot{y}=p_2/m$,
and assume a magnetic field of spatial dependence $F(x,y)=B\cdot e(y)$, with $e(y)$ the Heaviside step function,
$e(y)=1$ if $y\geq 0$, $e(y)=0$ if $y<0$, 
$\frac{de(y)}{dy}=\delta(y)$. B is constant. The set-up is sketched in figure 1a.

\setlength{\unitlength}{1cm}
\begin{picture}(12,5)
\put(1.7,0.7){figure 1a}
\put(.5,3){\line(1,0){4}}
\put(.8,4){{\large B}}
\put(1.5,2.6){\small{P}}
\put(3.1,2.6){\small{P'}}
\put(1.5,3){\circle*{.1}}
\put(3.5,3){\circle*{.1}}
\qbezier(1.5,3)(2.5,4)(3.5,3)
\put(1.5,3){\line(-1,-2){.8}}
\put(3.5,3){\line(3,-4){1}}
\put(0.3,2.2){$\vec{v}_0$}
\put(1.8,3.6){$\vec{v}'$}
\put(4.3,2.5){$\vec{v}''$}
\put(4.8,1.5){$\vec{v}$}
\thicklines
\put(1.5,3){\vector(1,1){.6}}
\put(3.5,3){\vector(1,-1){.6}}
\put(.75,2){\vector(1,2){.4}}
\put(4.09,2.2){\vector(3,-4){.5}}
\thinlines
\put(4,3.7){\vector(1,0){.7}}
\put(4,3.7){\vector(0,1){.7}}
\put(4.1,4.5){x}
\put(4.8,3.9){y}
\thinlines
\put(8.3,0.7){figure 1b}
\put(7.2,4){{\large B}}
\put(8,2.6){\small{P}}
\put(9.9,3.4){\small{P'}}
\put(8,3){\circle*{.1}}
\put(10,3.2){\circle*{.1}}
\put(6.8,2.2){$\vec{v}_0$}
\put(8.25,3.6){$\vec{v}'$}
\put(10.5,3){$\vec{v}''=\vec{v}$}
\put(7,3){\line(1,0){2.8}}
\put(9.8,3){\line(1,1){.5}}
\qbezier(8,3)(9,4)(10,3.2)
\put(8,3){\line(-1,-2){.8}}
\put(10,3.2){\line(1,-1){1}}
\thicklines
\put(8,3){\vector(1,1){.6}}
\put(10,3.2){\vector(1,-1){.6}}
\put(7.25,2){\vector(1,2){.4}}
\end{picture}

\noindent
At the point P a particle with initial velocity $\vec{v}_0$ 
passes from the lower half-plane (with vanishing magnetic field) into the upper half-plane, which has 
magnetic field $B$. 
At that same point the particle's velocity changes to $\vec{v}'$ if $\theta\neq 0$, as will be seen;
subsequently the usual commutative circular motion in a magnetic field $B$ follows. 
While exiting the superior half-plane at point P', the velocity will change to a final $\vec{v}$.  
Explicitely, Eq. (\ref{test1}) leads to 
\be
m\ddot{x}-F(y)\dot{y}=-m\theta B\delta(y)\dot{x}\dot{y}, \qquad
m\ddot{y}+F(y)\dot{x}=m\theta B\delta(y)\dot{x}^2.
\ee 
Using $\delta(y)=\delta(v_y t)=\frac{\delta(t)}{|v_y|}$,
one has
\be
m\ddot{x}-F(y)v_y=-m\theta B\delta(t)\dot{x}\frac{v_y}{|v_y|}, \qquad
m\ddot{y}+F(y)v_x=m\theta B\delta(t)\frac{v_x^2}{|v_y|}. \label{pre_shift}
\ee 
Integrating from $t=-\epsilon$ to $t=\epsilon$ (and taking $\epsilon\rightarrow 0$), one obtains
\be
v'_x=v_x(1-\theta B),\qquad v'_y=v_y(1+\theta B\frac{v^2_x}{v^2_y}).\label{shift}
\ee
As long as the velocity has a component orthogonal to the magnetic field gradient, i.e. if $v_x\neq 0$,
an instantaneous velocity change of order $\theta$ takes place at the point at which the particle enters the magnetic field. 
No instantaneous displacement appears. The change in kinetic energy is only of order
$\theta^2$, $(v'_x)^2+(v'_y)^2=v^2_x+v^2_y+\theta^2B^2(v^2_x+v^4_x/v^2_y)$.
The opposite mechanism works while the particle exits the magnetic field region.
This would render the correction quadratic in $\theta$, hence quite small.
To avoid that, figure 1b proposes a slightly modified configuration, which allows the particle
to exit the magnetic field along the gradient. In this case no unconventional velocity shift
occurs, and the particle keeps the $O(\theta)$ correction (\ref{shift}), with $v_x=v''_x=v'_x$, $v_y=v''_y=-v'_y$. 

It is important that the noncommutative correction (\ref{shift}) depends also on the ratio $\frac{v^2_x}{v^2_y}$;
this allows us to 
choose a favorable situation.
Assuming that experimental control can be maintained for velocity ratios
$\frac{v_x}{v_y}\sim 10^5$, a field $B\sim 1 T$, and considering a proton of mass $m_p$ and charge $1.6\times 10^{-19}C$,
one sees total reflection along the line PP' if at least
\be
\sqrt{\theta}\sim 2\times 10^{-13}meters. \label{bound1}
\ee 
Although this preliminary bound may  not seem a very strong one, 
compared to the $10^{-18}m$ presently reached in particle accelerators, it is remarkable
that a simple, classical, low-energy - hence cheap - set-up could achieve that.
Moreover, the bound can be significantly improved
by constructing a system of magnets designed to keep the proton under periodic motion.
Individually, each magnet will act on the charged particle as in figure 1b.
The velocity shifts will always add, rendering the bound cummulative. Then, $10^{2N}$ revolutions will
further decrease the bound on $\sqrt{\theta}$ by a factor of $10^{-N}$. One may also measure 
corrections of about one per cent in the velocity shift (\ref{shift}), and not of order one as total reflection required. 
The above considerations lead to a theoretical bound 
\be
\sqrt{\theta}\sim 2\times 10^{-14-N}meters, \label{bound2}
\ee 
where $10^{2N}$ is the number of revolutions performed by the particle inside the system of magnets. 
According to experience with the much higher energies used at particle accelerators, 
$N=5$ appears to be easily achievable, pushing the noncommutativity scale in the $TeV$ regime:
$\Lambda_{NC}\equiv\theta^{-1/2}\sim 1TeV$! This gives a bound similar to the much more complex best estimates performed
either via quantum mechanical \cite{bounds1} of quantum field theoretic \cite{bounds2} techniques. 
Stronger claims \cite{bounds3} are manifestly based on involved considerations of radiative corrections.
The bound (\ref{bound2}) will of course be affected by experimental errors or difficulties. 
A realistic set-up will most probably require numerical work in order to follow the trajectories in
a real magnetic field, which cannot be as simple as in figures 1a,b.

Although the above low-energy, cummulative, clasical experiment might seem attractive,
it is likely that the interference effects characteristic of quantum mechanics, associated with the choice
of a convenient set-up, will further improve the bound. This is currently under investigation.
The main purpose of this note has been to demonstrate a clear effect of classical noncommutativity
-- anomalous deflection in magnetic field gradients -- in as simple a system as possible.

\bigskip

{\bf Acknowledgements}

It is a pleasure to thank K.N. Anagnostopoulos, G.G. Athanasiu and T.N. Tomaras for useful discussions.
This project was partly supported by a European Community individual Marie Curie fellowship, 
under contract HPMF-CT-2000-1060.
The subsequent hospitality of Prof. Tomaras at Crete University is gratefully acknowledged.


\end{document}